\documentstyle[12pt]{article}

\catcode`\@=11

\newcount\hour
\newcount\minute
\newtoks\amorpm \hour=\time\divide\hour by 60\minute
=\time{\multiply\hour by 60 \global\advance\minute by-\hour}
\edef\standardtime{{\ifnum\hour<12 \global\amorpm={am}%
        \else\global\amorpm={pm}\advance\hour by-12 \fi
        \ifnum\hour=0 \hour=12 \fi
        \number\hour:\ifnum\minute<10
        0\fi\number\minute\the\amorpm}}
\edef\militarytime{\number\hour:\ifnum\minute<10
0\fi\number\minute}

\def\draftlabel#1{{\@bsphack\if@filesw {\let\thepage\relax
   \xdef\@gtempa{\write\@auxout{\string
      \newlabel{#1}{{\@currentlabel}{\thepage}}}}}\@gtempa
   \if@nobreak \ifvmode\nobreak\fi\fi\fi\@esphack}
        \gdef\@eqnlabel{#1}}
\def\@eqnlabel{}
\def\@vacuum{}
\def\marginnote#1{}
\def\draftmarginnote#1{\marginpar{\raggedright\scriptsize\tt#1}}
\def\draft{
        \pagestyle{plain}
        \overfullrule=2pt
        \oddsidemargin -.5truein
        \def\@oddhead{\sl \phantom{\today\quad\militarytime} \hfil
        \smash{\Large\sl DRAFT} \hfil \today\quad\militarytime}
        \let\@evenhead\@oddhead
        \let\label=\draftlabel
        \let\marginnote=\draftmarginnote
        \def\ps@empty{\let\@mkboth\@gobbletwo
        \def\@oddfoot{\hfil \smash{\Large\sl DRAFT} \hfil}
        \let\@evenfoot\@oddhead}
        \def\@eqnnum{(\theequation)\rlap{\kern\marginparsep\tt\@eqnlabel}%
        \global\let\@eqnlabel\@vacuum}  }

\newcommand{\rf}[1]{(\ref{#1})}
\renewcommand{\theequation}{\thesection.\arabic{equation}}
\renewcommand{\thefootnote}{\fnsymbol{footnote}}

\def\appendix#1{\addtocounter{section}{1}\setcounter{equation}{0}
\renewcommand{\thesection}{\Alph{section}}
\section*{Appendix \thesection\protect\indent \parbox[t]{11.15cm}{#1}}
\addcontentsline{toc}{section}{Appendix \thesection\ \ \ #1}}

\def\hi{{\hat{i}}}
\def\hj{{\hat{j}}}
\def\hk{{\hat{k}}}

\def\f{{\rm f}}

\def\cN{{\cal N}}

\def\alppr{{\alpha^\prime}}
\def\bepr{{\beta^\prime}}

\def\nline{\,\nabla\kern -0.7em\raise0.2ex\hbox{/}\,\,}
\def\yline{\,y\kern -0.47em /}
\def\aline{\,a\kern -0.49em /}
\def\parline{\,\partial\kern -0.55em /\,\,}

\def\bepr{{\beta^\prime}}

\def\rTr{{\rm Tr}}
\def\rTrpr{\hbox{\'Tr}}

\def\be{\begin{equation}}
\def\ee{\end{equation}}

\def\thzb{\theta^{\bar{Z}}}

\begin{document}

%\draft

\begin{flushright}
FIAN/TD/10-03 \\
hep-th/0301009
\end{flushright}

\vspace{1cm}

\begin{center}

{\Large \bf Superfield formulation of $\cN=4$ Super Yang-Mills

\bigskip
theory in plane wave
background\footnote{Contribution to Proceedings of the
3rd International Sakharov conference on physics.
June 24-29, Moscow, Russia.}}

\vspace{2.5cm}

R.R. Metsaev\footnote{
E-mail: metsaev@lpi.ru }

\vspace{1cm}

{\it Department of Theoretical Physics, P.N. Lebedev Physical
Institute, \\ Leninsky prospect 53,  Moscow 119991, Russia }

\vspace{3.5cm}

{\bf Abstract}

\end{center}

The superfield light cone gauge formulation of
$\cN =4$ SYM in plane wave background is developed. We find a
realization of superconformal symmetries in terms of a
light cone superfield. An oscillator realization of superconformal
symmetries is also obtained.

\newpage
\setcounter{page}{1}
\renewcommand{\thefootnote}{\arabic{footnote}}
\setcounter{footnote}{0}

%%%%%%%%%%%%%%%%%%%%%%%%%%%%%%%%%%%%%%%%%%%
\section{Introduction}
%%%%%%%%%%%%%%%%%%%%%%%%%%%%%%

The new maximally supersymmetric solution of IIB
supergravity with Ramond-Ramond flux \cite{blau} attracted recently
considerable interest. On the one hand,
the light cone gauge Green-Schwarz superstring
action in this background is quadratic in both  bosonic and
fermionic superstring $2d$ fields, and therefore, this model can
be explicitly quantized \cite{rrm0112}. On the other hand,  in
\cite{beren} it was proposed that this superstring in plane wave
background corresponds to a certain (large R-charge)  sector of
the $\cN=4$ SYM theory. Given that the plane wave superstring model
can be quantized explicitly \cite{rrm0112,rrm0202} one can study
the duality correspondence between string states and gauge theory
operators at the string-mode level \cite{beren}. The  new duality
\cite{beren} turned out
to be very fruitful and this renewed interest in various aspects of
string/gauge theory correspondence.

One of the important aspects of the string/gauge theory correspondence
is to find explicit and precise relation between string
states and appropriate operators of gauge theory.
For the case of plane wave duality solution to this problem was
suggested in \cite{beren} by introducing the operators which are
often referred to as BMN operators.
These operators were extensively studied in the literature
(see for instance
\cite{Chu:2002qj,Beisert:2002bb,Constable:2002vq}
and references in \cite{Beisert:2002tn})
and various
improvements of the original BMN operators were
suggested \cite{Kristjansen:2002bb,Bianchi:2002rw,Parnachev:2002kk,
Beisert:2002tn,Minahan:2002ve}.
Unfortunately,
the guiding principles which would allow one to fix the precise
form of these operators are not clear so far\footnote{It seems
natural to expect that these operators should be derivable from the
microscopic open-string
approaches to the supersymmetric plane wave $Dp$ branes.
Discussion $Dp$ branes in the framework of
microscopic approach may be found in \cite{Chu:2002in,skenderis,
Billo:2002ff,Bak:2002rq,Bergman:2002hv}.}.
Recent interesting discussion of this theme may be found in
\cite{Corley:2002ny}. One of the important inputs in establishing
a precise
correspondence might be the geometry of space-time in which fields
of SYM are propagating. Usually, the Hamiltonian
formulation of conformal field theory for SYM is
considered
in $R\times S^3$ background\footnote{Discussion of
SYM theory in various curved backgrounds may be found in
\cite{Blau:2000xg}. Plane wave SYM does not fall in the cases
considered in \cite{Blau:2000xg}. Detailed study of the
Hamiltonian formulation of SYM in $R\times S^3$ background may be
found in \cite{oku}. Relevance of a curved background
for SYM was discussed (in the original AdS/CFT context) in
\cite{Park:2000du}.}. The space-time symmetries of this
background, however, are in sharp contradiction with the (super)symmetries
of the plane wave superstring model.
Therefore, taking into account that
some (super)symmetries of plane wave SYM action
are look the same as (super)symmetries of the type IIB superstring in
plane wave Ramond-Ramond background it was suggested
\cite{Metsaev:2002sg} that the SYM theory
in plane wave background may
be preferable in the context of the new duality.

Another argument in favor of the discussion of the plane wave SYM is that
$4d$ plane wave background can be obtained via Penrose
limit from $R\times S^3$ background. Given that the SYM in $R\times S^3$
is used for study of holography in the original context of AdS/CFT
correspondence we expect
that   $4d$ plane wave background is most appropriate for the study
of holographical issues of the plane wave duality.
Thus, it seems natural to expect that in order to
match the states of the plane wave superstring and operators of SYM
the Penrose limit on the
string theory side should be supplemented by the Penrose limit
on the SYM side.
Various alternative discussions of
plane wave holography may be found in \cite{Das:2002cw}.

In \cite{Metsaev:2002sg} we developed the
Hamiltonian light cone gauge formalism
for the plane wave SYM in component form. In this paper we continue the
study of the
plane wave SYM and develop a superfield formulation of this theory.
Superfield formulation turns out to be more compact, and we expect
that this formalism will be helpful in future studies (similar to one
in \cite{Santambrogio:2002sb}).
Independently of the above string/gauge theory duality motivation
it seems interesting to study SYM theory in curved backgrounds and the
present work may be also considered as an effort in this direction.

\section{Action of plane wave $\cN=4$ SYM in light cone superspace}

Component form of light cone gauge action of $\cN=4$ SYM is
formulated in terms bosonic fields $A^I=(A^\hi,\phi^M)$, $A^\hi$
is spin one field and $\phi^M$ are six scalar fields, and
$16$-component fermionic field $\psi^\oplus$ which satisfies the
constraint $\bar\gamma^-\psi^\oplus=0$. Because on plane wave
Ramond-Ramond superstring theory side $so(8)$ symmetries are
broken to $so(4)\oplus so'(4)$ it is natural to reduce manifest
SYM R-symmetries $so(6)$ to the R-symmetries $so(2)\oplus so'(4)$.
Another technical advantages in reducing manifest SYM R-symmetries
is related with the fact in this way one can formulate SYM theory
in terms of unconstrained light cone superfield. As in the flat
space-time the superfield formulation of $\cN=4$ SYM could be
developed by using a certain superspace whose 16 real-valued
fermionic coordinates related with the $\theta^\alpha$ which form
Glifford algebra \be \{\theta^\alpha,\theta^\beta\}
=\frac{1}{4}(\gamma^+)^{\alpha\beta}\,, \qquad
\bar\gamma^+\theta=0\,. \ee However the superfield based on such
fermionic coordinates should satisfy complicated constraints
\cite{Brink:1983pf}. To avoid such superfield description one
needs to introduce Grassmann coordinates instead of Clifford
coordinates $\theta^\alpha$ \cite{Brink:1982pd,Green:fu}. To this
end we break manifest R-symmetries $so(6)$ to R-symmetries
$so(2)\oplus so'(4)$ and introduce the fermionic `coordinates'
$\theta^Z$ and fermionic `momenta' $\theta^{\bar{Z}}$ by relations
\be \theta^Z = \frac{1}{2}\gamma^{\bar{Z}}\bar\gamma^Z\theta\,,
\qquad \theta^{\bar{Z}} =
\frac{1}{2}\gamma^Z\bar\gamma^{\bar{Z}}\theta\,, \qquad \gamma^Z =
\frac{1}{\sqrt{2}}(\gamma^3+{\rm i}\gamma^4)\,,\quad
\gamma^{\bar{Z}} = \gamma^{Z*}\,, \ee which satisfy the desired
commutation relations \be
\{\theta^{Z\alpha},\theta^{\bar{Z}\beta}\}
=\frac{1}{8}(\gamma^+\gamma^{\bar{Z}} \gamma^Z)^{\alpha\beta}\,,
\qquad \{\theta^{Z\alpha},\theta^{Z\beta}\}=0\,, \qquad
\{\theta^{\bar{Z}\alpha},\theta^{\bar{Z}\beta}\}=0\,. \ee
Accordingly the fields of SYM are also decomposed into multiplets
of R-symmetry $so(2)\oplus so'(4)$ algebra. The six scalar fields
$\phi^M$ are decomposed into four scalar fields $\phi^{i'}$
transforming in vector representation of the $so'(4)$ algebra and
two complex fields $Z$, $\bar{Z}$ defined by \be Z \equiv
\frac{1}{\sqrt{2}}(\phi^3+{\rm i}\phi^4)\,, \qquad
\bar{Z}\equiv\frac{1}{\sqrt{2}}(\phi^3 - {\rm i}\phi^4)\,, \ee
while the fermionic field $\psi^\oplus$ is decomposed as follows
\be \psi^\oplus = \psi^{\oplus Z}+\psi^{\oplus \bar{Z}}\,,\qquad
\qquad \psi^{\oplus
Z}\equiv\frac{1}{2}\gamma^{\bar{Z}}\bar\gamma^Z \psi^\oplus\,,
\qquad \psi^{\oplus\bar{Z}} \equiv \frac{1}{2}\gamma^Z
\bar\gamma^{\bar{Z}} \psi^{\oplus}\,. \ee

Light-cone gauge $\cN=4$ SYM in plane wave background
can be formulated in
light-cone superspace which
is based on space-time coordinates $x^\pm$, $x^\hi$
and Grassmann coordinates $\theta^Z$
\footnote{$\mu=0,1,\ldots 9$ are $so(9,1)$ vector indices;
$I,J,K=1,\ldots 8$ are
$so(8)$ transverse indices; $\hi,\hj,\hk=1,2$ are space-time symmetry
$so(2)$ transverse
indices; $M,N=3,\ldots, 8$ are R-symmetry $so(6)$ indices;
$i',j'=5,6,7,8$ are R-symmetry $so'(4)$ indices; indices
$\hat{I},\hat{J}=1,2,5,6,7,8$ are collection of space-time $so(2)$
indices $\hi,\hj$ and R-symmetry $so'(4)$ indices $i',j'$.
Coordinates in light-cone directions are defined by
$x^\pm \equiv (x^9\pm x^0)/\sqrt{2}$. Remaining transverse coordinates
$x^I$ are decomposed into $x^\hi$, $x^{i'}$, $x^{Z,\bar{Z}}$ where
$x^{Z,\bar{Z}}\equiv (x^3\pm {\rm i}x^4)/\sqrt{2}$.
The scalar product of two  $so(8)$ vectors is decomposed then as
$X^I Y^I = X^\hi Y^\hi + X^{i'}Y^{i'}+ X^ZY^{\bar{Z}} + X^{\bar{Z}} Y^Z$.}.
In this light-cone superspace we introduce scalar
superfield $\phi(x^\pm,x^\hi,\theta^Z)$. Instead of position space it is
convenient
to use momentum space for all coordinates except the light-cone time  $x^+$.
This implies using momenta $p^+$, $p^\hi$,
$\theta^{\bar{Z}}$, instead of positions $x^-$, $x^\hi$, $\theta^Z$
respectively.
Thus we consider the scalar superfield
$\phi(x^+, p^+,p^\hi,\theta^{\bar{Z}})$ with the
following expansion in powers of Grassmann momentum $\theta^{\bar{Z}}$
\begin{eqnarray}
\phi(p,\thzb) & = & p^+\bar{Z} +
{\rm i}\thzb\bar\gamma^{\bar{Z}}\psi^{\oplus\bar{Z}}
-\frac{1}{2}\thzb\bar\gamma^{-\bar{Z}\hi}\thzb A^{\hi}
-\frac{1}{2}\thzb\bar\gamma^{-\bar{Z}i'}\thzb \phi^{i'}
\label{supfie}\nonumber\\
&-&
\frac{\rm i}{6p^+}\thzb\bar\gamma^{-\bar{Z}\hat{I}}
\thzb \thzb\bar\gamma^{\hat{I}}\psi^{\oplus Z}
-\frac{1}{24p^+}(\thzb\bar\gamma^{-\bar{Z}\hat{I}}\thzb)^2Z\,.
\end{eqnarray}
The fact that the fields $A^I(p)$, $\psi^{\oplus}(p)$ are subject
to the hermitian conjugation rules given in \rf{fieherrul}
can be expressed in terms of the superfield by the following relation
\be
\phi(-p,\thzb) = (p^+)^2\int
d^4\theta^{\bar{Z}\dagger} e^{2\thzb \bar\gamma^- \thzb{}^\dagger/p^+}
\phi(p,\thzb)^\dagger\,.
\ee

In terms of the superfield $\phi(p,\theta^{\bar{Z}})$ the action
of $\cN=4$ SYM takes then the form \be S = S_2 +S_3 +S_4\,, \ee
where the free action $S_2$ and cubic action $S_3$ are fixed to
be\footnote{Because the expression for $S_4$ is not illuminating
we do not present it here.}
\begin{eqnarray}
\label{s2sup}
&&
S_2 = \rTr
\int d^4\theta^{\bar{Z}} dx^+d^3p\, p^+\phi(-p,-\thzb)
({\rm i}\partial_{x^+}
+P^-)\phi(p,\thzb)\,,
\\
\label{s3sup}
&&
S_3 =\rTr\,\,\, \frac{1}{3}\int d\Gamma_3\,\,
{\bf P}^\hi\frac{\Theta^{\bar{Z}}
\bar\gamma^{-\bar{Z}\hi}\Theta^{\bar{Z}}}{p_1^+p_2^+p_3^+}
\prod_{a=1}^3\phi(p_a,\theta^{\bar{Z}}_a)\,.
\end{eqnarray}
The expression for the particle Hamiltonian $P^-$ which enters the
free action is given in \rf{pppm}, while the integration measure
in cubic action is given by
\be
d\Gamma_3  = \frac{1}{(2\pi)^{3/2}}
\delta^{(3)}(\sum_{a=1}^3p_a)
\delta(\sum_{a=1}^3\theta_a^{\bar{Z}})
\prod_{a=1}^3 d^3p_a
d^4\theta_a^{\bar{Z}}\,.
\ee
The quantities ${\bf P}^\hi$, $\Theta^{\bar{Z}}$ are defined by
\be\label{pgen}
{\bf P}^\hi \equiv \frac{1}{3} \sum_{a=1}^3 p_a^\hi(p_{a+1}^+-p^+_{a+2})\,,
\qquad
\Theta^{\bar{Z}}
 \equiv \frac{1}{3}\sum_{a=1}^3 \theta^{\bar{Z}}_a(p^+_{a+1}-p^+_{a+2})\,,
\ee
so that they are manifestly symmetric under cyclical permutations
of particle labels 1,2,3. Taking into  account the momenta conservation
laws we obtain the relations
\be
{\bf P}^\hi = {\bf P}_{12}^\hi={\bf P}_{23}^\hi={\bf P}_{31}^\hi\,,
\qquad
\Theta^{\bar{Z}}
 = \Theta_{12}^{\bar{Z}}=\Theta_{23}^{\bar{Z}}=\Theta_{31}^{\bar{Z}}\,,
\ee
where
\be
{\bf P}_{ab}^\hi\equiv p_a^\hi p_b^+ - p_b^\hi p_a^+\,,
\qquad
\Theta_{ab}^{\bar{Z}}\equiv \theta^{\bar{Z}}_a p_b^+
- \theta_b^{\bar{Z}} p_a^+\,.
\ee
The superfield form of the action \rf{s2sup},\rf{s3sup}
was obtained by direct
comparison with the action taken in terms of component fields
\rf{kinpw},\rf{lag3}.

In the Hamiltonian approach the physical fields satisfy the
canonical (anti)commu\-tation relations
\be
[A^I(p)^{\sf a},A^J(p')^{\sf a'}]|_{equal\,\, x^+}
=\frac{1}{2p^+}\delta^{IJ}\delta^{(3)}(p+p')\hbox{I}{}^{\,\sf a\, a'}\,,
\ee
\be
\{\psi^{\oplus Z\alpha}(p)^{\sf a},
\psi^{\oplus \bar{Z}\beta}(p')^{\sf a'}\}|_{equal\,\, x^+}
=-\frac{1}{4}(\gamma^-\bar\gamma^{\bar{Z}}\gamma^Z)^{\alpha\beta}
\delta^{(3)}(p+p')\hbox{I}{}^{\,\sf a\, a'}\,,
\ee
where $\hbox{I}{}^{\,\sf a\, a'}$ is a projector
operator which we insert to respect the Lie algebra indices of the
physical fields $A^I=A^I{}^{\sf a}t_{\sf a}$ and
$\psi^\oplus=\psi^{\oplus}{}^{\sf a}t_{\sf a}$. All that is required this
operator should satisfy the relation
$\hbox{I}{}^{\,\sf a\, c}\, \rTr (t_{\sf c} t_{\sf b}) =
\delta_{\sf b}^{\sf a}$. Taking into account the expression for
$\delta$-function over anti-commuting variables \rf{delfun}
the above-given canonical commutation relations
can be collected into the superfield form
\be
[\phi(p,\theta^{\bar{Z}})^{\sf a},
\phi(p{}',\,\theta^{\bar{Z}}{}')^{\sf a'}]|_{equal\,\, x^+}
=\frac{1}{2p^+}\delta^{(3)}(p+p')\delta(\theta^{\bar{Z}}
+\theta^{\bar{Z}}{}')\hbox{I}{}^{\,\sf a\, a'}\,.
\ee

\section{Superfield realization of
superconformal symmetries of $\cN=4$ plane wave SYM
}\label{GLOSYM}

In this section we discuss realization of global
superconformal supersymmetries of $\cN=4$ SYM
in plane wave background.
As is well known these symmetries are generated by
the $psu(2,2|4)$ superalgebra.
The realization of these symmetries on the component fields of $\cN=4$ SYM
was obtained in \cite{Metsaev:2002sg}. Here we
develop a superfield realization of
these symmetries.
To do that we use the
framework of Noether charges. The Noether charges play an
important role in analysis of the symmetries of dynamical systems.
The choice of the light cone gauge spoils manifest global
symmetries,  and  in order to demonstrate that these global
invariances are still present, one needs to  find the Noether
charges that generate them.

The component
form of the Noether charges for $\cN=4$ SYM was obtained in
\cite{Metsaev:2002sg}.
Superfield representation for the
Noether charges can be derived by direct comparison
with the Noether charges taken in the component form.
Doing this we obtain the following representation for the
Noether charges $G_{f.t.}$
\be
G_{f.t.} = \int d^4\theta^{\bar{Z}}
d^3p\,\, p^+ \phi(-p,-\theta^{\bar{Z}})G
\phi(p,\theta^{\bar{Z}})\,,
\ee
where $G$ denotes representation of generators of the
$psu(2,2|4)$ superalgebra in terms of differential operators
acting on the superfield $\phi(p,\theta^{\bar{Z}})$.
These differential operators are listed below.

{\bf Generators of the isometry symmetries}:
\begin{eqnarray}
\label{pppm}&&
P^+ = p^+\,,\qquad
P^- = -\frac{p^2}{2p^+} + \frac{\f^2}{2}p^+\partial_p^2\,,
\\
&&
T^\hi = e^{-{\rm i}\f x^+}(p^\hi -\f p^+\partial_{p^\hi})\,,
\qquad
\bar{T}^\hi = e^{{\rm i}\f x^+}(p^\hi + \f p^+\partial_{p^\hi})\,,
\\
&&
J^{\hi\hj}= l^{\hi\hj}
+M^{\hi\hj}\,,\qquad
l^{\hi\hj}\equiv p^\hi\partial_{p^\hj}-p^\hj\partial_{p^\hi}
\qquad M^{\hi\hj}\equiv \theta^Z\bar\gamma^{-\hi\hj}\thzb\,.
\end{eqnarray}
where we use notation $p^2=p^\hi p^\hi$, $\partial_p^2
=\partial_{p^\hi}\partial_{p^\hi}$.
These generators satisfy commutation relations of the
isometry symmetries algebra of $4d$ plane wave background
(see Appendix C).

{\bf Proper conformal generators}:
\begin{eqnarray}
D & = & -2\partial_{p^+}p^+ -\partial_{p^\hi}p^\hi
+ 2\theta^Z\bar\gamma^-\theta^{\bar{Z}}-1\,,
\\
C & = &e^{-2{\rm i}\f x^+}(
\frac{p^2}{2p^+}
+\frac{\f^2}{2}p^+\partial_p^2
-\f \partial_{p^\hi} p^\hi
+\f)\,,
\\
C^\hi&=& e^{-{\rm i}\f x^+}\Bigl(
-\partial_{p^+}p^\hi
-\frac{1}{2p^+}\partial_{p^\hi}p^2
+M^{\hi\hj}\frac{p^\hj}{p^+}
+(\theta^Z\bar\gamma^-\theta^{\bar{Z}}-1)\frac{p^\hi}{p^+}
\nonumber\\
& + & \f( - \frac{1}{2}\partial_p^2 p^\hi
+\partial_{p^\hi}
(\partial_{p^+}p^+ +\partial_{p^\hi}p^\hi
-\theta^Z\bar\gamma^-\theta^{\bar{Z}})
- M^{\hi\hj}\partial_{p^\hj}\Bigr)\,,
\\
K^-&=& \frac{1}{4p^+}\partial_p^2p^2+
(\partial_{p^+} +\frac{1}{p^+}(1-\theta^Z\bar\gamma^-\theta^{\bar{Z}}))
(\partial_{p^+}p^+ +\partial_{p^\hi}p^\hi
-\theta^Z\bar\gamma^-\theta^{\bar{Z}})
\nonumber\\
&+ & \frac{1}{2p^+}M^{\hi\hj}l^{\hi\hj}
+ \frac{1}{2p^+}M^{\hi\hj}M^{\hi\hj}\,.
\end{eqnarray}
Generators of the isometry symmetries and proper conformal symmetries
form commutation relations of the $so(4,2)$ algebra taken to be in
the plane wave basis.

{\bf Generators of R-symmetries so(6) algebra}:
\begin{eqnarray}
&& J^{i'j'}= \theta^Z\bar\gamma^{-i'j'}\theta^{\bar{Z}}\,,
\qquad\quad
J^{Z\bar{Z}}=\theta^Z\bar\gamma^-\theta^{\bar{Z}}-1\,,
\\[6pt]
&&
J^{Zi'}
= \frac{p^+}{2}\theta^Z\bar\gamma^{-Zi'}\theta^Z\,,
\qquad
J^{\bar{Z}i'}
= \frac{1}{2p^+}\thzb\bar\gamma^{-\bar{Z}i'}\thzb\,.
\end{eqnarray}
The hermitian properties of these generators are fixed to be
$J^{i'j'\dagger}
= - J^{i'j'}$,
$J^{Zi'\dagger}= - J^{\bar{Z}i'}$, $J^{Z\bar{Z}\dagger}
= J^{Z\bar{Z}}$.

{\bf Supercharges}:
\begin{eqnarray}
\label{omp}&&
\Omega^+ =2\bar\gamma^-(p^+\theta^Z+\theta^{\bar{Z}})\,,
\\
&&
\Omega^-=2T^\hi\bar\gamma^\hi
(\theta^Z + \frac{1}{p^+}\theta^{\bar{Z}})\,,
\qquad
\bar\Omega{}^-
=2\bar{T}{}^\hi\bar\gamma^\hi(\theta^Z+
\frac{1}{p^+}\theta^{\bar{Z}})\,,
\end{eqnarray}
\be
\label{smin}S^- = 2(D-1)(\theta^Z+\frac{1}{p^+}\theta^{\bar{Z}})
+(l^{\hi\hj}+2M^{\hi\hj})
\gamma^{\hi\hj}(\theta^Z + \frac{1}{p^+}\theta^{\bar{Z}})\,.
\ee
The hermitian properties of the supercharges are fixed to be
$\Omega^{+\dagger}=\Omega^+$, $\Omega^{-\dagger}=\bar\Omega^-$,
$S^{-\dagger}=-S^-$.

Transformations of fields defined by
$\delta_G\phi= [\phi,G_{f.t.}]|_{eq.\,\, x^+}$ take
then the form
\be
\delta_G \phi = G\phi\,.
\ee

\section{Oscillator realization of superconformal \\
symmetries}

The expressions for generators above-given are defined in terms of
differential operators acting on wave function taken to be in momentum
space. The representation for generators can be simplified by passing to
the oscillators representation of wave function.

Let $\phi(p,p^+)$ be square integrable solution to free equations
of motion for fields of plane wave $\cN=4$ SYM corresponding to
the $p^+>0$, (dependence on $\theta^{\bar{Z}}$ is implicit)
\be\label{equmot0} {\rm i}\partial_{x^+} \phi =- P^-\phi\,,\qquad
p^+>0\,. \ee Transformation of this wave function from momentum
space to the oscillator basis wave function denoted by
$|\phi(a,p^+)\rangle$ is fixed to be \be |\phi(a,p^+)\rangle =
\int d^2p\,\, e^F \phi(p,p^+)|0\rangle\,, \qquad F \equiv
-\frac{1}{2\f p^+}p^\hi p^\hi +\sqrt{\frac{2}{\f p^+}}\, a^\hi
p^\hi -\frac{1}{2} a^\hi a^\hi\,, \ee $\f>0$, where $a^\hi$ is
creation operator and $|0\rangle$ is the Fock space vacuum defined
by $\bar{a}^\hi|0\rangle=0$, $[\bar{a}^\hi,a^j]=\delta^{\hi\hj}$.
These formulas imply the following realization of operators
$a^\hi$ and $\bar{a}^\hi$ \be\label{app} a^\hi =
\frac{1}{\sqrt{2\f p^+}}\, (p^\hi - \f p^+\partial_{p^\hi})\,,
\qquad \bar{a}^\hi = \frac{1}{\sqrt{2\f p^+}}\,( p^\hi + \f
p^+\partial_{p^\hi})\,. \ee Making use of these relation it is
easy to see that the solution to the equation \rf{equmot0} can be
presented in the following elegant form \be |\phi(a,p^+)\rangle =
\sum_{n=0}^\infty e^{-{\rm i}\f(n+1)x^+} a^{\hi_1}\ldots a^{\hi_n}
\phi^{\hi_1\ldots \hi_n}(p^+)|0\rangle\,. \ee Taking into account
a realization of the differential operator $\partial_{p^+}$ on the
oscillator wave function $|\phi(a,p^+)\rangle$ \be
\Bigl(\partial_{p^+} +\frac{1}{4p^+}(a^\hi a^\hi -\bar{a}^\hi
\bar{a}^\hi -2)\Bigr)|\phi(a,p^+)\rangle = \int d^2p\,\, e^F
\partial_{p^+}\phi(p,p^+) \ee and the formulas \rf{app} it is easy
to transform the generators to the oscillator basis. For the
generators of the isometry symmetries we get the representation
\begin{eqnarray}
&&
P^+ =p^+\,, \qquad P^- = -\f  (a^\hi\bar{a}^\hi+1)\,,
\\
&&
T^\hi =\sqrt{2\f\,p^+}\,e^{-{\rm i}\f x^+} a^\hi\,,
\qquad \bar{T}^\hi
=\sqrt{2\f\,p^+}\, e^{{\rm i}\f x^+} \bar{a}{}^\hi\,,
\\
&&
J^{\hi\hj} = a^\hi\bar{a}^\hj -a^\hj\bar{a}^\hi
+\theta^Z\bar\gamma^{-\hi\hj}\theta^{\bar{Z}}\,.
\end{eqnarray}

Oscillator representation for generators of the proper conformal
transformations takes the form
\begin{eqnarray}
&&
D= -2\partial_{p^+}p^+ + 2\theta^Z\bar\gamma^-\theta^{\bar{Z}}-1\,,
\\
&&
C=\f e^{-2{\rm i}\f x^+} a^\hi a^\hi\,,
\\
&&
C^\hi = -\partial_{p^+}T^\hi -\frac{1}{2\f p^+}
(\bar{T}^\hi C +T^\hi P^-)+\frac{1}{p^+}\theta^Z
\bar\gamma^-\gamma^\hi
\bar\gamma^\hj\theta^{\bar{Z}}T^\hj\,,
\\
&&
K^-= \frac{1}{8\f^2p^+}\{C,\bar{C}\}
-\frac{1}{4\f^2p^+}\, (P^-)^2
+(\partial_{p^+} -\frac{1}{p^+}\theta^Z\bar\gamma^-\theta^{\bar{Z}})
(\partial_{p^+}p^+ +1 -\theta^Z\bar\gamma^-\theta^{\bar{Z}})
\nonumber
\\
&& \hspace{1cm}+ \frac{1}{p^+}M^{\hi\hj}a^\hi \bar{a}^\hj
+\frac{1}{2p^+}M^{\hi\hj}M^{\hi\hj} +\frac{3}{4p^+}\,.
\end{eqnarray}
The oscillator representation for the supercharges is obtainable
by inserting the above-given expressions for bosonic generators
into formulas \rf{omp}-\rf{smin}. Because of certain attractive
features it seems that the oscillators representation is fruitful
direction to go. Interesting discussion of application of the
oscillator construction to the plane wave physics
may be found in \cite{Fernando:2002wv}\footnote{Realization
of conformal and Heisenberg algebras in plane wave CFT
correspondence may be found in \cite{Das:2002ij}.}.

{\bf Acknowledgments}.
This work was supported by the INTAS project 991590,
by the RFBR Grant 02-02-17067 and RFBR Grant
for Leading Scientific Schools 00-15-96566.

\bigskip
{\bf Appendix A: Lagrangian of plane wave $\cN=4$ SYM
in terms of component fields on space-time}.
Light cone gauge fixed Lagrangian can be presented as follows
\be {\cal L}= {\cal L}_2+{\cal L}_3+{\cal L}_4\,, \ee where ${\cal
L}_2$ describes quadratic part of Lagrangian while ${\cal L}_3$
and ${\cal L}_4$ describe 3-point and 4-point interaction vertices
respectively. The quadratic part ${\cal L}_2$ is given by
\be\label{kinpw} {\cal L}_2 =\rTrpr\,\, \frac{1}{2}A^I\Box A^I
-\frac{\rm i}{4}\psi^\oplus \frac{\bar{\gamma}^+}{\partial^+}\Box
\psi^\oplus\,, \ee
where the standard covariant D'Alembertian operator in
$4d$ plane wave background
\be\label{4dpw} ds^2 = 2dx^+ dx^- -\f^2 x^\hi x^\hi  dx^+dx^+ +
dx^\hi dx^\hi\,, \qquad \hi=1,2,\ee
is given by
\be\label{box} \Box = 2\partial_{x^+}\partial_{x^-} +
\partial_{x^\hi}\partial_{x^\hi}
+ \f^2 x^\hi x^\hi\partial_{x^-}\partial_{x^-}\,. \ee The
3-point and 4-point interaction vertices are given by
\begin{eqnarray}\label{lag3}
{\cal L}_3 & = &\rTrpr\,\,
 -[A^I,A^J]\partial^I  A^J -
\partial^J A^J\frac{1}{\partial^+} [A^I,\partial^+A^I]
\\
& + &{\rm i} \partial^IA^I \frac{1}{\partial^+}
(\psi^\oplus\bar{\gamma}^+\psi^\oplus ) +\frac{\rm
i}{4}[A^I,\psi^\oplus]
\frac{\bar{\gamma}^+\gamma^I\bar{\gamma}^J}{\partial^+}
\partial^J\psi^\oplus
+ \frac{\rm i}{4}\partial^J\psi^\oplus
\frac{\bar{\gamma}^+\gamma^J\bar{\gamma}^I}{\partial^+}
[A^I,\psi^\oplus]\,, \nonumber
\\
&&\nonumber
\\
\label{lag4} {\cal L}_4 & =  &\rTrpr\,\, -\frac{1}{4}[A^I,A^J]^2
-\frac{1}{2} (\frac{1}{\partial^+}[A^J,\partial^+ A^J])^2
\nonumber\\
& + & \frac{{\rm
i}}{\partial^+}(\psi^\oplus\bar{\gamma}^+\psi^\oplus)
\frac{1}{\partial^+}[A^I,\partial^+ A^I] +\frac{\rm
i}{4}[A^I,\psi^\oplus]\frac{\bar{\gamma}^+\gamma^I
\bar{\gamma}^J}{\partial^+}
[A^J,\psi^\oplus]
\nonumber\\
& + &
\frac{1}{2}(\frac{1}{\partial^+}(\psi^\oplus\bar{\gamma}^+
\psi^\oplus))^2\,, \qquad\partial^I=\partial_{x^I}\,,\quad
\partial^+=\partial_{x^-}\,,\quad
\partial^-=\partial_{x^+}\,.
\end{eqnarray}
The physical fields $A^I$, $\psi^\oplus$ being Lie algebra valued
in the adjoint representation of gauge group are assumed to be
anti-hermitian $(A^I)^\dagger =-A^I$,
$\psi^{\oplus\dagger}= - \psi^\oplus$.
The $\rTrpr$ denotes minus trace $\rTrpr
Y \equiv  - \rTr Y$. All fields are assumed to be independent
of six coordinates $x^M$: $\partial_{x^M} A^I =0$,
$\partial_{x^M} \psi^\oplus =0$.

{\bf Appendix B: Component form of the action of $N=4$ SYM in
terms of momentum modes}. Because a formulation of action in terms of
momentum modes of fields is more convenient in
superfield approach we make the Fouirer transformation
\be
\phi(x) = {\rm i}\int \frac{d^3p}{(2\pi)^{3/2}}
e^{i(p^+x^- +x^\hi p^\hi)}\phi(x^+,p)\,,
\qquad d^3p \equiv dp^1 dp^2 dp^+\,,
\ee
so that the Fouirer modes satisfy the hermitian conjugation rules
\be\label{fieherrul}
A^I(x^+,p)^\dagger = A^I(x^+,-p)\,,
\qquad
\psi^\oplus(x^+,p)^\dagger = \psi^\oplus(x^+,-p)\,.
\ee
In terms of momentum modes the quadratic and the cubic parts of
the action \rf{kinpw},\rf{lag3} take the form
\begin{eqnarray}
&&
S_2 = \rTr\,\int\! dx^+ d^3p\, A^I(-p)p^+({\rm i}\partial^- +P^-)A^I(p)
-\frac{\rm i}{2}\psi^\oplus(-p)\bar\gamma^+
({\rm i}\partial^- +P^-)\psi^\oplus(p)\,,\,\,\,\,\,\,\,\,\,\,\,\,\,\,
\\
&&
S_3 = \rTr\int d\Gamma_3(p)\Bigl(
\frac{2}{p_3^+}{\bf P}^K \delta^{IJ}
A^I(p_1)A^J(p_2)A^K(p_3)
\nonumber\\
&& \hspace{1cm} -\frac{p_1^+ - p_2^+}{2p_1^+p_2^+p_3^+}{\bf P}^I
\psi(p_1)\bar\gamma^+\psi(p_2) A^I(p_3)
-\frac{1}{2p_1^+p_2^+}{\bf P}^I
\psi(p_1)\bar\gamma^{+IJ}\psi(p_2) A^J(p_3)\Bigr)\,,
\end{eqnarray}
where the momentum ${\bf P}^I$ is defined in \rf{pgen},
while the measure $d\Gamma_3(p)$ is given by
\be
d\Gamma_3(p)  = \frac{1}{(2\pi)^{3/2}}
\delta^{(3)}(\sum_{a=1}^3p_a)
\prod_{a=1}^3 d^3p_a\,.
\ee
In these formulas we
keep dependence on six coordinates $x^M$. To get expressions
corresponding to $4d$ SYM one needs to apply rule ${\bf P}^M=0$.

{\bf Appendix C: Commutation relations of the $psu(2,2|4)$
superalgebra in plane wave basis}. These commutation relations
for anti-hermitian bosonic generators were worked out in
\cite{Metsaev:2002sg}. In this paper we use the basis in which some of
the generators have different hermitian properties. This is to say that
$P^\pm$, $K^-$ are taken to be hermitian, while the hermitian properties of
the remaining generators are fixed to be
$T^{\hi\dagger}=\bar{T}^\hi$, $J^{\hi\hj\dagger}=-J^{\hi\hj}$,
$D^\dagger =-D$,
$C^\dagger = \bar{C}$,
$C^{\hi\dagger}=-C^\hi$\footnote{Denoting by $G_{ah}$
the bosonic anti-hermitian generators used in
\cite{Metsaev:2002sg} we have the following relations:
$P^\pm ={-\rm i}P^\pm_{ah}$, $T^\hi= -{\rm i}T_{ah}^\hi$,
$\bar{T}^\hi= -{\rm i}\bar{T}_{ah}^\hi$,
$J^{\hi\hj}=J_{ah}^{\hi\hj}$,
$D=D_{ah}$,
$C={\rm i}C_{ah}$, $\bar{C}={\rm i}\bar{C}_{ah}$,
$C^\hi=C_{ah}^\hi$,
$\bar{C}^\hi= \bar{C}_{ah}^\hi$,
$K^-={\rm i}K^-_{ah}$. Here we use the same supercharges $\Omega^\pm$ as in
\cite{Metsaev:2002sg},
while $S^-$ is multiplied by factor ${\rm i}$.}.

In plane wave basis all
generators, by definition, are eigenvectors of $P^-$ and $D$.
Therefore given generator
$G$ we introduce $q_G(P^-)$ and $q_G(D)$ charges which are
defined by commutators
\be\label{pdcha} [P^-, G] = -\f q_G(P^-)G\,, \qquad [D,G] =
q_G(D)G\,. \ee These charges for generators of the $psu(2,2|4)$
superalgebra are given in Table.

\begin{center}
\begin{tabular}{|c|c|c|c|c|c|c|c|c|c|c|c|c|}
\hline        && && && && && &&
\\ [-3mm]\ \ \  Generators  && && && && && &&
\\
& $P^+$ & $T^\hi$ & $\bar{T}^\hi$  & $\!K^-$&  $C^\hi$ &
$\bar{C}^\hi$ & $\!C$ & $\bar{C}$ & $\Omega^- $ & $\bar{\Omega}^-$
& $\Omega^+$ &  $\!S^-$
\\
$ q $ - charges \ \ \ \ \ \  &&  && &&   && && &&
\\ \hline
&& && &&    && &&  &&
\\[-3mm]
$q_G(P^-)$ charge&0& $1$ &$-1$&   $0$&1&$-1$&2& $-2$ & 1& $-1$& 0&
0
\\[2mm]\hline
&&&&&&&&&& &&
\\[-3mm]
$q_G(D)$ charge   & $-2$ & $-1$ & $- 1$ & 2 & 1& 1 & 0& 0 & 0& 0 &
$-1$ &  1
\\[2mm]\hline
\end{tabular}
\end{center}
Note that the $q$-charges of the generators $J^{\hi\hj}$ and
$J^{MN}$ are equal to zero. Remaining (anti)commutators can be
collected in several groups.

Commutators between elements of isometry algebra are
\be\label{pwcomrel1} [\bar{T}^\hi,T^\hj]=2\f \delta^{\hi\hj}
P^+\,,\qquad [T^\hi,J^{\hj\hk}]=\delta^{\hi\hj}T^\hk
-\delta^{\hi\hk}T^\hj\,.\ee

Commutators between proper conformal generators are given by
\be\label{pwcomrel2} [\bar{C}^\hi,C^\hj]=2\delta^{\hi\hj}K^-\,,
\qquad [\bar{C},C]=-4\f P^-\,,
\qquad [\bar{C}^i,C]=2\f C^i\,. \ee

Commutators between generators of isometry algebra and proper
conformal generators take the form \be\label{pwcomrel3}
[P^+,K^-]=D\,, \qquad [P^+,C^\hi]= T^\hi\,, \qquad
[T^\hi,K^-]=C^\hi\,, \ee \be\label{pwcomrel4}
[T^\hi,C^\hj]=\delta^{\hi\hj}C\,, \qquad [T^\hi,\bar{C}]= -2
\f \bar{T}^\hi\,, \ee \be\label{pwcomrel5} [\bar{C}^\hi,T^\hj] =
\delta^{\hi\hj}(P^- + \f D) + 2\f J^{\hi\hj}\,. \ee
Commutators between isometry algebra and supercharges are
\be
[T^\hi,\bar{\Omega}{}^-] = \f \bar\gamma^{+\hi}\Omega^+\,, \qquad
[T^\hi,S^-]=\gamma^\hi \Omega^-\,, \qquad [P^+,S^-] =
 \gamma^+
\Omega^+\,, \ee
$$
[\Omega^\pm,
J^{\hi\hj}]=\frac{1}{2}\bar\gamma^{\hi\hj}\Omega^\pm\,, \qquad
[S^-, J^{\hi\hj}]=\frac{1}{2}\gamma^{\hi\hj}S^-\,.
$$

Commutators between proper conformal generators and supercharges
are \be [K^-,\Omega^+] = -\frac{1}{2}\bar\gamma^- S^-\,, \qquad
[C^\hi,\Omega^+] = -\frac{1}{2}\bar\gamma^{-i} \Omega^-\,, \qquad
[C^\hi,\bar{\Omega}^-] = -\f \bar\gamma^\hi S^-\,, \ee \be
[C,\bar{\Omega}^-] = -2\f\, \Omega^-\,. \ee

Commutators of $so(6)$ algebra
and commutators between generators of the $so(6)$ and supercharges
in plane wave basis take the standard form.

Anticommutators between supercharges are \be
\{\Omega^+,\Omega^+\}=2\bar{\gamma}^-P^+\,,\quad
\{\Omega^-,\Omega^-\}=-2\bar{\gamma}^+C\,,
\quad
\{S^-,S^-\}=4\gamma^+K^-\,,
\ee \be
\{\bar{\Omega}^-,\Omega^-\}= 2\bar\gamma^+ P^-
+\f\bar\gamma^{+\hi\hj}J^{\hi\hj} - \f \bar\gamma^{+MN}J^{MN}\,,
\ee \be \{\Omega^-,\Omega^+\} =
\bar\gamma^+\gamma^-\bar\gamma^\hi T^\hi\,, \qquad
\{\Omega^-,S^-\} = - 2 \bar\gamma^{+\hi} C^\hi\,, \ee \be
\{S^-,\Omega^+\} =\gamma^+\bar{\gamma}^-D +\frac{1
}{2}\gamma^+\bar{\gamma}^-\gamma^{\hi\hj}J^{\hi\hj} - \frac{1
}{2}\gamma^+\bar{\gamma}^-\gamma^{MN}J^{MN}\,. \ee Modulo
(anti)commutators obtainable from the ones above-given by applying
hermitian conjugation, the remaining (anti)commutators are equal
to zero.

{\bf Appendix D: Fierz identities}. Often we use the
following identities for
$\gamma$-matrices\footnote{$m,n=0,1,2,9$ are so(3,1) indices;
$M,N=3,\ldots,8$ are R-symmetry so(6) indices;
$\alpha,\beta,\gamma,\delta=1,\ldots,16$ are spinor  indices. We use
convention for $\gamma$ matrices following Ref. \cite{Metsaev:2002sg}.}
\be\label{basfieide}
 \gamma^\mu_{\alpha(\beta}\gamma^\mu_{\delta\gamma)}=0\,,
\qquad\delta^\alpha_\alppr \delta^\beta_\bepr -
\delta^\alpha_\bepr \delta^\beta_\alppr
 =\frac{1}{48}(\gamma^{\mu\nu\rho})^{\alpha\beta}
(\gamma^{\mu\nu\rho})_{\alppr\bepr}\,, \ee
\be\label{unf}
\delta^\alpha_\beta \delta^\delta_\gamma
-\frac{1}{2}(\gamma^{mn})^\alpha{}_\beta (\gamma^{mn})^\delta{}_\gamma
+\frac{1}{2}(\gamma^{MN})^\alpha{}_\beta (\gamma^{MN})^\delta{}_\gamma
+(\beta\leftrightarrow \gamma)
=2\gamma^m_{\beta\gamma}(\gamma^m)^{\alpha\delta}\,.
\ee
The identities \rf{basfieide} are well known, while
the identity \rf{unf} is derivable from the 1st identity in \rf{basfieide}.
From the 2nd identity \rf{basfieide} we can get the following
`generating' identity
\be\label{basfie}
8\psi\bar\gamma^-\theta^Z  \theta^Z\bar\gamma^-\varphi =
\psi\bar\gamma^{-\bar{Z}\hat{I}}\varphi
\theta^Z\bar\gamma^{-Z\hat{I}}\theta^Z\,.
\ee
Inserting in \rf{basfie}
$\psi = \theta^Z\gamma^{Z\hat{I}}$,
$\varphi = \gamma^{Z\hat{J}} \theta^Z$
and $\psi = \theta^Z\gamma^{Zi'}$, $\varphi = \gamma^{+Z}\psi^{\oplus Z}$
we get the following respective identities

\be\label{fieide11} \theta^Z\bar\gamma^{-Z\hat{I}}\theta^Z
\theta^Z\bar\gamma^{-Z\hat{J}}\theta^Z =
\frac{1}{6}\delta^{\hat{I}\hat{J}}
\theta^Z\bar\gamma^{-Z\hat{K}}\theta^Z
\theta^Z\bar\gamma^{-Z\hat{K}}\theta^Z\,, \ee \be
\theta^Z\bar\gamma^{-Zi'}\theta^Z \theta^Z\bar\gamma^Z\psi^{\oplus
Z} = -\frac{1}{6}\theta^Z\bar\gamma^{-Z\hat{I}}\theta^Z
\theta^Z\bar\gamma^{\hat{I}}\gamma^{Zi'}\psi^{\oplus Z}\,. \ee
Multiplying the identity \rf{unf} by $\gamma^+$, $\gamma^-$ and
then by $\theta^Z$, $\theta^Z$ gives the identity \be
(\bar\gamma^\hi\theta^Z)_\alpha(\bar\gamma^\hi\theta^Z)_\beta
=-\frac{1}{8}(\bar\gamma^{+\bar{Z}i'})_{\alpha\beta}
\theta^Z\bar\gamma^{-Zi'}\theta^Z\,. \ee

{\bf Appendix E: Integrals over anti-commuting variables}. The
measure $d^4\theta^Z$ is normalized to be \be\label{basint} \int
d^4\theta^Z\,\, \theta^{Z\alpha } \theta^{Z\beta
}\theta^{Z\gamma}\theta^{Z \delta }
=\frac{1}{64}(\gamma^{+\bar{Z}\hat{I}})^{\alpha\beta}
(\gamma^{+\bar{Z}\hat{I}})^{\gamma\delta}\,. \ee Note that the
l.h.s. of this formula is antisymmetric in spinor indices
$\alpha$, $\beta$, $\gamma$, $\delta$. Making use of the 1st
identity \rf{basfieide} one can make sure that the r.h.s is also
antisymmetric in $\alpha$, $\beta$, $\gamma$, $\delta$. The fact
that the integral \rf{basint} can be cast into the form
above-given can be proved by using identity \be \theta^{Z\alpha}
\theta^{Z\beta} =
\frac{1}{16}(\gamma^{+\bar{Z}\hat{I}})^{\alpha\beta}
\theta^Z\bar\gamma^{-Z\hat{I}}\theta^Z\,, \ee and identity
\rf{fieide11}. All remaining integrals are derivable from the
basic integral \rf{basint} in a straightforward way:
\begin{eqnarray}
&&
\int d^4\theta^Z\,\, (\theta^Z \bar\gamma^-\theta^{Z\dagger})^4
=\frac{1}{16}(\theta^{Z\dagger}
\bar\gamma^{-\bar{Z}\hat{I}}\theta^{Z\dagger})^2\,,
\\
&&
\int d^4\theta^Z\,\, \theta^Z\bar\gamma^{-Z\hat{I}}\theta^Z\,\,
\theta^Z\bar\gamma^{-Z\hat{J}}\theta^Z
=4\delta^{\hat{I}\hat{J}}\,,
\\
&&
\int d^4\theta^Z\,\,
\theta^{Z\alpha }\theta^{Z\beta }\,\,\theta^Z
\bar\gamma^{-Z\hat{I}}\theta^Z
=\frac{1}{4}(\gamma^{+\bar{Z}\hat{I}})^{\alpha\beta}\,.
\end{eqnarray}
Taking these relations into account the expression for
$\delta$-function is fixed to be
\be\label{delfun}
\delta(\theta^Z)=
\frac{1}{24}
\theta^Z\bar\gamma^{-Z\hat{I}}\theta^Z
\theta^Z\bar\gamma^{-Z\hat{I}}\theta^Z\,.
\ee

\newpage
%%%%%%%%%%%%%%%%%%%%%%%%%%%%%%%

\end{document}